\documentclass[ aps,%
floatfix,%
final,%
notitlepage,%
oneside,%
onecolumn,%
nobibnotes,%
nofootinbib,%
superscriptaddress,%
showpacs,%
]%
{revtex4}

\usepackage{epsfig}
\usepackage{amsfonts}
\usepackage{amsmath}
\usepackage{epsfig}
\usepackage{graphics}

\newcommand{\JP}{\psi} \newcommand{\ec}{\eta_c}

\newcommand{\epem}{e^+e^-}

\newcommand{\beq}{\begin{eqnarray}}\newcommand{\eeq}{\end{eqnarray}}
\newcommand{\beqa}{\begin{eqnarray*}}\newcommand{\eeqa}{\end{eqnarray*}}

\begin{document}

\title{Double charmonium production at B-factories within light cone formalism.}
\author{V.V. Braguta}
\email{braguta@mail.ru}
\affiliation{Institute for High Energy Physics, Protvino, Russia}

\begin{abstract}
This paper is devoted to the study of the processes 
$e^+e^- \to  J/\Psi \eta_c, J/\Psi \eta_c', \psi' \eta_c, \psi' \eta_c'$ within light cone formalism. 
It is shown that if one disregards the contribution of higher fock 
states, the twist-3 distribution amplitudes needed in the calculation
can be unambiguously determined from the twist-2 distribution amplitudes and equations 
of motion. Using  models of the twist-2 distribution amplitudes 
the cross sections of the processes under study have been calculated. 
The results of the calculation are in agreement with Belle and BaBar experiments. 
It is also shown that relativistic and radiative corrections to the cross sections
play crucial role in the achievement of the agreement between the theory and experiments. 
The comparison of the results of this paper with the results obtained in other 
papers has been carried out. In particular, it is shown that the results of papers where 
relativistic and radiative corrections were calculated within NRQCD are overestimated by
a factor of $\sim 1.5$.
\end{abstract}


\maketitle

\newcommand{\ins}[1]{\underline{#1}}
\newcommand{\subs}[2]{\underline{#2}}
%%%%%%%%%%%%%%%%%%%%%%%%%%%%%%%%%%%%%%%%%%%%%%%
\vspace*{-1.cm}
\section{Introduction}

Double charmonium production at B-factories is very interesting problem 
from theoretical point of view. At the beginning, the interest to this 
problem was caused by large discrepancy between theoretical predictions for the cross section 
of the process $e^+e^- \to J/\Psi \eta_c$ \cite{Braaten:2002fi, Liu:1, Liu:2} and it's 
first experimental measurement by Belle collaboration \cite{Abe:2002rb}. Lately, large discrepancy 
between theory and experiment was found by Belle \cite{Abe:2004ww} and BaBar \cite{Aubert:2005tj} 
collaborations for the processes $e^+e^- \to J/\Psi \eta_c', \psi' \eta_c, \psi' \eta_c', 
J/\Psi \chi_{c0}, \psi' \chi_{c0}$.

A number of attempts were made to explain this discrepancy. For instance, the authors of papers 
\cite{Bodwin:2002fk, Bodwin:2002kk, Luchinsky:2003yh} studied the possibility of admixture 
of $\epem\to J/\psi J/\psi $ events in the  $e^+e^- \to J/\Psi \eta_c$. 
Another attempt was to attribute the discrepancy to large radiative corrections 
\cite{Zhang:2005ch, Gong:2007db, Zhang:2008gp}. One more  very popular 
approach to the problem \cite{Ma:2004qf, Bondar:2004sv, Braguta:2005gw, 
Braguta:2005kr, Braguta:2006nf, Ebert:2006xq, Choi:2007ze, Berezhnoy:2007sp, Ebert:2008kj}
consisted in  taking into account internal motion of the quark-antiquark pair in charmonium.  
Some other interesting points of view on this problem were proposed in papers \cite{Lee:2003db, Zhang:2008ab, 
Berezhnoy:2006mz}.

Among many approaches to the resolution of this challenging problem one should mention 
nonrelativistic QCD (NRQCD) \cite{Bodwin:1994jh}. Contrary to many other models 
of quarkonium production this approach allows one to improve systematically the accuracy 
of the calculation. Thus at the leading order approximation \cite{Braaten:2002fi, Liu:1, Liu:2} there 
is very large discrepancy between NRQCD predictions for the cross section of the process $e^+e^- \to J/\Psi \eta_c$ 
and the experiments. However, after the inclusion of the one loop radiative corrections \cite{Zhang:2005ch, Gong:2007db}
this discrepancy becomes smaller. At the next step, after taking into account radiative and relativistic 
corrections simultaneously \cite{He:2007te, Bodwin:2007ga}, the problem can be resolved. From 
this example one can draw a conclusion, that relativistic and radiative corrections play 
very important role in the processes with charmonia production. Probably, in a similar way 
one can resolve many other puzzles with quarkonia production \cite{Bodwin}. 

It should be noted that within NRQCD the discrepancy between the theory and experiments was resolved 
only for the process $e^+e^- \to J/\Psi \eta_c$. At the same time the discrepancies 
were observed also for the processes $e^+e^- \to J/\Psi \eta_c', \psi' \eta_c, \psi' \eta_c', J/\Psi \chi_{c0}, \psi' \chi_{c0}$. 
It is not clear is it possible to apply the same approach for these processes, since excited 
charmonia states are rather relativistic and the 
application of NRQCD to the production of such states is questionable. 

Another systematic approach to the study of hard exclusive processes is light cone formalism (LC) 
\cite{Lepage:1980fj, Chernyak:1983ej}. Within this approach the amplitude of hard exclusive process
can be separated into two parts. The first part is partons production at very small 
distances, which can be treated within perturbative QCD. The second part is 
the hadronization of the partons at larger distances. This part contains  information about 
nonperturbative dynamic of the strong interactions. For hard exclusive processes it can be 
parameterized by process independent distribution amplitudes (DA), which can be considered as hadrons' wave functions
at light like separation between the partons in the hadron. It should be noted that within LC 
one does not assume that the mesons are nonrelativistic. This approach can  equally well be
 applied to the production of light and heavy mesons, if the DAs of the produced meson are known. 
For this reason, one can hope that within this approach one can study the production of excited charmonia states. 

The first attempts to describe the experimental results obtained at Belle and BaBar collaborations within LC were done in papers 
\cite{Ma:2004qf, Bondar:2004sv, Braguta:2005kr, Braguta:2006nf}. There are two very important problems common for 
all these papers. The first one is that in these papers the renormalization 
group evolution of the DAs was disregarded. What is very important since the evolution of the DAs takes into account 
very important part of radiative corrections -- the leading logarithmic radiative corrections. 
The second problem is poor knowledge of charmonia DAs, which are the key ingredient of 
any calculation done within LC. 

Recently, the leading twist DAs of the $S$-wave charmonia 
mesons have become the objects of intensive study \cite{Choi:2007ze, Bodwin:2006dm, Ma:2006hc, Braguta:2006wr, 
Braguta:2007fh, Braguta:2007tq, Feldmann:2007id, Bell:2008er, Hwang:2008qi}.
Knowledge about these DAs allowed one to build some models for the $S$-wave charmonia DAs, 
that can be used in practical calculations. In this paper LC will be applied to the study
the processes $e^+e^- \to J/\Psi \eta_c, J/\Psi \eta_c', \psi' \eta_c, \psi' \eta_c'$. 
It will be shown that with the models of DAs proposed in papers \cite{Braguta:2006wr, 
Braguta:2007fh, Braguta:2007tq} LC predictions are  in  agreement with 
the results obtained at Belle and BaBar experiments. 
 
This paper is organized as follows. In the next sections a brief description 
of LC is given. In section III the formula for the amplitude of the 
process $e^+ e^- \to V P$, where $V$ and $P$ are vector and pseudoscalar mesons, is derived. In section IV 
this formula is used to calculate the cross sections of the processes 
$e^+e^- \to J/\Psi \eta_c, J/\Psi \eta_c', \psi' \eta_c, \psi' \eta_c'$. 
The results of the calculation are discussed in section V. Finally, in the last section 
the results of this paper are summarized.

\section{Brief description of light cone formalism.}

In this section a brief description of light cone formalism (LC) will be given. As an example, let us 
consider  hard exclusive process with single meson production. And let us assume that this 
meson is a pseudoscalar meson $P$. The presence of high energy scale $E_h$, which is of order of the 
characteristic energy of the hard exclusive process, allows one to apply factorization theorem for the
amplitude of the process $T$ 
\beq
T= \sum_{n} C_n \times \langle P | O_n | 0 \rangle, 
\label{fact}
\eeq
where the coefficient $C_n$ describes partons production at small distances, 
the matrix element $\langle P | O_n | 0 \rangle$ describes hadronization of the partons which takes 
place at large distances. The sum is taken over all possible operators $O_n$. For instance, 
the operators $\bar Q \gamma_{\mu} \gamma_5 Q, 
\bar Q \gamma_{\mu} \gamma_5 {\overset {\leftrightarrow} { D}_{\mu_1}} {\overset {\leftrightarrow} { D} }_{\mu_2} Q, 
\bar Q \sigma_{\mu \nu} \gamma_5 G_{\alpha \beta} Q$ are few examples of the operator $O_n$. 
Actually, there are infinite 
number of the operators $O_n$ that contribute to the pseudoscalar meson production. 

The cross section of hard exclusive process can be expanded in  inverse powers of the high
energy scale $E_h$
\beq
\sigma = \frac {\sigma_0} { E_h^n} + \frac {\sigma_1} { E_h^{n+1}} + ... 
\label{exp}
\eeq
To determine if some operator contributes to a given term in $1/E_h$ expansion 
one uses the concept of the twist of this operator \cite{Braun:2003rp}. Thus only  the leading twist 
-- the twist-2 operators $\bar Q \hat z \gamma_5 Q, \bar Q \hat z \gamma_5 ({z \overset {\leftrightarrow} { D} })Q, 
\bar Q \hat z \gamma_5 ({z \overset {\leftrightarrow} { D} })^2 Q, ...$\footnote{$z$ here is lightlike fourvector $z^2=0$.}
contribute to the leading term in expansion (\ref{exp}). 
From this one sees that already at the leading order approximation infinite number of operators 
contribute to the cross section. Nevertheless, it is possible to cope with infinite 
number of contributions if one parameterizes all  the twist-2 operators by the moments of some function $\phi(x)$
\beq
\langle P(q) | \bar Q \hat z \gamma_5 
({-i \overset {\leftrightarrow} { D} }_{\mu_1})... ({-i \overset {\leftrightarrow} { D} }_{\mu_n}) Q  |0 \rangle \times z^{\mu_1}...z^{\mu_n}  = 
i f_P (qz)^{n+1} \int_{0}^1  dx \phi(x) (2x-1)^n,
\eeq
where $q$ is the momentum of the pseudoscalar meson $P$, $f_P$ is the constant which is defined as
$\langle P(q) | \bar Q \gamma_{\mu} \gamma_5 Q |0 \rangle = i f_P q_{\mu}$, $x$ is the 
fraction of momentum of the whole meson $P$ carried by quark. The function $\phi(x)$ is called the leading 
twist distribution amplitude (DA). One can think of it as about the Fourier transform of the  wave function 
of the meson $P$ with lightlike distance between quarks.  Using the definition of this function, factorization theorem (\ref{fact})
can be rewritten as
\beq
T=\int_0^1 dx H(x) \phi(x),
\label{amp}
\eeq
where $H(x)$ is the hard part of the amplitude, which describes small distance effects.
This part of the amplitude can be calculated within perturbative QCD. As it was 
noted, the leading twist DA parameterizes infinite set of the twist-2 operators. 
This part of the amplitude describes hadronization of quark-antiquark pair at large distances
and parameterizes nonperturbative effects in the amplitude. It should be noted 
that formula (\ref{amp}) resums the contributions of all the twist-2 operators. If the meson $P$ 
is a nonrelativistic meson, formula (\ref{amp}) resums relativistic corrections to the amplitude $T$. 

Now let us consider radiative corrections to formula (\ref{amp}). The presence of two different 
energy scales, which are strongly separated $E_h \gg M_P$, gives rise to the appearance of large logarithm
$\log {E_h^2/M_P^2}$. This logarithm enhances the role of radiative corrections. The main contribution 
to amplitude (\ref{amp}) comes from the leading logarithmic radiative corrections 
$\sim (\alpha_s \log {E_h^2/M_P^2})^n$. It turns out that these corrections can be taken 
into the account in formula (\ref{amp}) as follows \cite{Chernyak:1983ej, Lepage:1980fj}
\beq
T=\int_0^1 dx H(x, \mu) \phi(x, \mu).
\label{amp1}
\eeq
To resum the leading logarithmic corrections coming from all loops the scale $\mu$ should
be taken of order of $\sim E_h$. The hard part of the amplitude $H(x, \mu)$ should be calculated 
at the tree level approximation. At this level $H(x, \mu)$ depends on the renormalization scale $\mu$ only through 
the running of the strong coupling constant $\alpha_s(\mu)$.
The rest of the leading logarithms are resummed in the DA $\phi(x, \mu)$ using renormalization group 
method (see Appendix B). It should be stressed that formula (\ref{amp1}) exactly resums the leading logarithmic 
radiative corrections. 

Commonly, to study the production of nonrelativistic mesons one uses 
effective theory NRQCD \cite{Bodwin:1994jh}. NRQCD deals with three energy scales $m_Q \gg m_Q v \gg m_Q v^2$,
where $m_Q$ is the mass of the heavy quark $Q$, $v\ll 1$ is relative velocity of quark antiquark pair. 
In the process of hard nonrelativistic meson production there appears one additional energy scale $E_h$
which is much greater than all scales $m_Q, m_Q v, m_Q v^2$. Evidently, it is not 
possible to apply NRQCD at this scale. From the effective theory perspective,
first, this large energy scale must be integrated out. And this is done through 
the taking into account renormalization group evolution of the DA $\phi(x, \mu)$.

This paper is devoted to the study of the process $e^+e^- \to V  P$, where $V$ and $P$ are  
vector and pseudoscalar mesons. Essential 
feature of this process is that for it the leading order contribution in $1/E_h$ 
expansion is zero \cite{{Bondar:2004sv}}. So, we deal with the next-to-leading twist process. 
For this process formula (\ref{amp}) remains valid. The only difference is that 
now we have contributions coming from different twist-2 and twist-3 DAs (see next section). 
As in the case of the leading twist process, formula (\ref{amp}) resums relativistic 
corrections to the amplitude. However, if we consider the leading logarithmic 
radiative corrections to the amplitude, formula (\ref{amp1}) is incorrect. 
One can expect that some leading logarithms are lost. It is only possible to state
that formula (\ref{amp1}) 
resums the leading logarithms which appear in the amplitude due to the 
running of the $\alpha_s$ and DAs. Below this approximation will be used. 

There is one common feature of all next-to-leading twist process. 
It is connected to the following fact: along with two particles 
twist-3 operators (for instance $\bar Q \gamma_5 Q$)
 there appears operators of the type $\bar Q \gamma_{\mu} \gamma_5 G_{\alpha \beta} Q$.
Evidently, these operators describe higher fock state $| \bar Q Q g >$ contribution 
to the amplitude of the process. NRQCD predicts that for nonrelativistic mesons such states 
are suppressed by higher powers of relative velocity of quark-aniquark pair inside the meson 
\cite{Bodwin:1994jh}. For this reason, in this paper the contribution 
of such states will be disregarded.

\section{The amplitude of the process: $e^+e^- \to V  P$.}

\begin{figure}[t]
\begin{center}
\includegraphics[scale=0.8]{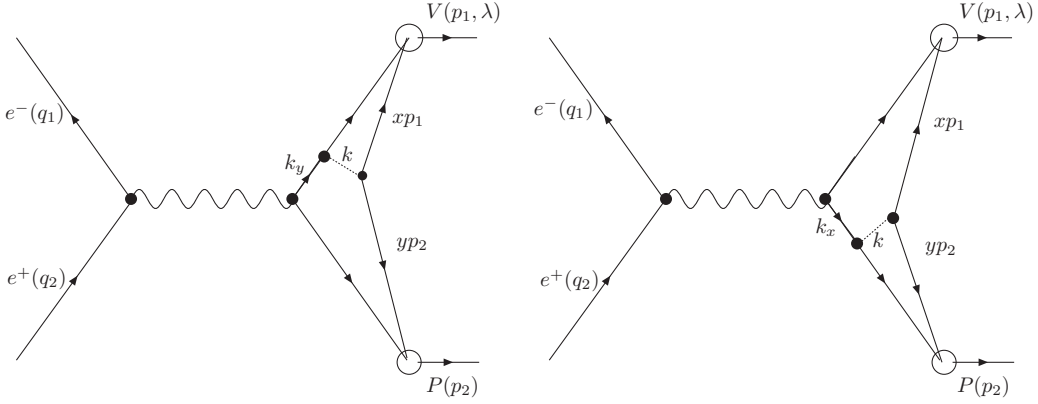} 
\caption{The diagrams that contribute to the process $e^+e^- \to V (p_1,\lambda) P(p_2)$
at the leading order approximation in the strong coupling constant.}
\label{tr}
\end{center}
\end{figure}

In this section the amplitude of the process $e^+e^- \to V P$, where $V=J/\Psi, \psi'$ and $P=\eta_c, \eta_c'$ 
will be considered. Two diagrams that give contribution to the amplitude of this process  are 
shown in Fig  \ref{tr}. The other two can be obtained from the depicted ones by the charge conjugation. 
 The amplitude of the process involved can be written 
in the following form: 
\beq
M=-4 \pi \alpha \frac { \bar u(k_1) \gamma_{\mu} u(k_2) } {s} \langle V(p_1,\lambda) P(p_2) | J_{\mu}^{em} | 0 \rangle,
\eeq
where $\alpha$ is the electromagnetic coupling constant, $\bar u(k_1), u(k_2)$ are the electron and positron bispinors, 
$\sqrt s$ is the invariant mass of $e^+ e^-$ system, $J_{\mu}^{em}$ is the electromagnetic current. The matrix element 
$\langle V(p_1,\lambda) P(p_2) | J_{\mu}^{em} | 0 \rangle$ can be parameterized by the only formfactor $F(s)$:
\beq
\langle V(p_1,\lambda) P(p_2) | J_{\mu}^{em} | 0 \rangle = i q_c~ 
F(s)~ e_{\mu \nu \rho \sigma} \epsilon_{\lambda}^{\nu} p_1^{\rho} p_2^{\sigma},
\eeq
where $q_c$ is the charge of $c$ quark, $\epsilon_{\lambda}^{\nu}$ is the polarization vector
of the meson $V(p_1, \lambda)$.
The cross section of the process under consideration can be written as follows
\beq
\sigma (\epem\to\ V P) & = & \frac {\pi \alpha^2 q_c^2} {6}  \biggl ( \frac {2 |\mathbf{p}|} {\sqrt s}  \biggr )^3 |F(s)|^2.
\eeq
In the last formula $\mathbf{p}$ ~is the momentum of the meson $V$ in the center mass
frame of the final mesons.

To calculate the formfactor $F(s)$ LC will be applied. 
As it was noted in the previous section, within LC the formfactor is a series in inverse 
powers of the characteristic energy of the process
$\sqrt s$. The leading order contribution to the formfactor $F(s)$ in the $1/s$ expansion was 
obtained in paper \cite{Bondar:2004sv}. 
In derivation of this expression the authors of paper \cite{Bondar:2004sv} disregarded
the mass difference of the final mesons. For the mesons with different 
masses the expression for the formfactor $F(s)$, which  was derived in paper \cite{Braguta:2005kr}, 
can be written as follows
\beq
|F (s)| & = & \frac{32\pi}{9}   \left|\frac{f_V f_P  M_P M_V}{q_0^4}\right|\,I_0\,,
\label{previous}
\eeq
\beq
\label{I0}
I_0 & = &
  \int^1_0 dx \int^1_0 dy~ \alpha_s(\mu) \left\{
    \frac {M_P} {M_V^2} \frac{Z_t Z_p 
    V_{T}(x, \mu) P_{P}(y, \mu)}{d(x,y)\, s(x)}- \frac 1 {M_P} \frac{\overline {M}_c^2 }{{ M_V}^2}\,
    \frac{Z_m Z_t V_T(x, \mu) P_A(y, \mu)}{d(x,y)\,s(x)}+
\right.\\ \nonumber & + &
\frac{1}{2 M_P}\frac{V_{L}(x, \mu)\,P_{A}(y, \mu)}{d(x,y)}+\frac{1}{2 M_P}
\frac{(1-2 y)}{s(y)}\frac{V_{\perp}(x, \mu)\,P_{A}(y, \mu)}{d(x,y)}+
\\ & + & \left.
\frac{1}{8} \biggl ( 1-Z_t Z_m \frac{4{ \overline M}_c^2}{{ M_V}^2 }\biggr ) \frac 1 M_P \, 
\frac{(1+y)V_A(x, \mu)P_A(y, \mu)}
{d^2(x,y)}\right\}.
\eeq
Where $q_0^2 \simeq (s-M_V^2-M_P^2)$, $x$ and $y$ are the fractions of momenta carried by 
quark in the meson V and by quark in the meson P correspondingly, $P_A, P_P, V_T, V_L, V_{\perp}, V_A$ are 
the DAs defined in Appendix B, $M_V, M_P$ are the masses of the 
vector and pseudoscalar mesons correspondingly, $\overline {M}_c= M^{\overline {MS} }_c ( \mu = M^{\overline {MS}}_c )$, 
$Z_{t}$ and $Z_{p}$ are the renormalization factors of the local tensor and pseudoscalar currents, 
the dimensionless propagators $d(x,y), s(x), s(y)$ are defined as follows:
\beq
\label{prop}
d(x,y)& = & 
  \frac{k^2}{q_0^2}=\left( x+\frac{\delta}{y}\right)
  \left(y+\frac{\delta}{x}\right),~
s(x) = \frac {k_x^2} {q_0^2} - \delta = \left(x+\frac{\delta}{y (1-y) } \right),
~ s(y)= \frac {k_y^2} {q_0^2} - \delta = \left(y+\frac{\delta}{x (1-x) }
\right), \\ \nonumber
Z_p & = & \left[\frac{\alpha_s(\mu)}{\alpha_s({\overline M}_c^2)} \right]^{\frac{-3c_F}{b_o}},\quad
 Z_{t}=\left[\frac{\alpha_s(\mu)}{\alpha_s({\overline M}_c^2)} \right]^{\frac{c_F}{b_o}},\quad
 Z_{m}=\left[\frac{\alpha_s(\mu)}{\alpha_s({\overline M}_c^2)} \right]^{\frac{3c_F}{b_o}}, \quad 
 \delta=\Biggl (Z_m \frac{{\overline  M}_c}{q_0}\Biggr )^2\,, 
\eeq
where $c_F=4/3,\,b_o=25/3$, the definitions of the fourvectors $k, k_x$ and $k_y$ are 
shown in Fig. \ref{tr}. 

It should be noted here that if one ignores renormalization group running, 
formula (\ref{previous}) is the analog of formula (\ref{amp}). So, in this 
case, it takes into account infinite series of the relativistic corrections to double charmonium
 production. If one takes into account the leading logarithmic corrections 
due to the DAs and the running of the strong coupling constant, the scale 
$\mu$ should be taken of order of the characteristic energy of the 
process $\sim \sqrt s$.

Before we proceed with the calculation of double charmonium production,  the expression for the formfactor $F(s)$ 
will be modified as follows:

1. Formula (\ref{previous}) was obtained at the first nonvanishing approximation in the $1/s$ expansion. This implies that 
all parameters in this formula must be taken at the same level of accuracy. For instance, in the expression 
$q_0^2 \simeq (s-M_V^2-M_P^2)$ last two terms are beyond the accuracy of calculation and these terms must
be omitted. Below the following approximation will be used $q_0^2=s$. 

2. Now let us consider expressions (\ref{prop}) for the propagators $d(x,y), s(x), s(y)$. It is seen 
that all these expressions contain the terms proportional to $\sim \delta$. These terms are of NLO approximation, 
so, according to the previous item, they must be omitted. Thus the propagators can be written as
\beq
d(x,y)=x y, \quad s(x) = x, \quad s(y) = y.  
\label{prop1}
\eeq
However, if one substitutes these expressions to (\ref{I0}) and takes into account the renormalization group 
evolution of the DAs,  the divergence at the end point region ($x,y \sim 0$) appears.  
It should be noted that the motion of 
quark-antiquark pair in the end point regions $x,y \sim 0,1$ is relativistic. 
So, if we consider the production of a nonrelativistic meson without the evolution of the DAs,
the end point regions in the DA are strongly suppressed. For this reason 
the expression for the formfactor $F(s)$ is free from the divergence. However, if 
the evolution of the DAs is taken into account, the divergence in the $F(s)$ appears.
In this paper this problem will be solved as follows. According to the definition of $x$
\beq
x= \frac {E^c + p^c_z} {E^M+p^M_z}, 
\eeq
where $E^c, p^c_z$ are the energy and $z$ component of the momenum of  $c$-quark, $E^M, p^M_z$ are the energy and 
$z$ component of the momenum of meson $M$. If the energy and momenum of 
$c$-quark is much greater than it's mass, one can use the  propagators in  form (\ref{prop1}), 
since corrections to this approximation are suppressed.
Let us now consider the kinematic region where $c$-quark is approximately at rest and $x$ can be estimated as
\beq
x \sim \frac {M^*_c} {E^M+p^M_z} \simeq \frac {M_c^*} {\sqrt s}=x_{min}.
\label{xmin}
\eeq
$M_c^*$ here is the pole mass of $c$ quark. In this region the accuracy of  
propagators (\ref{prop1}) is not sufficient and one must take into account the 
corrections. It is clear that in any approach the corrections to the propagators regularize
the whole expression for the formfactor $F(s)$. This fact can be seen as follows:
for any kinematical region of double quark and double antiquark production shown 
in Fig. 1 the squares of momenta of the gluon and quark propagators cannot be smaller 
than $(2 M_c^*)^2$ and $(M_c^*+M_{V,P})^2$ correspondingly. So, there is no divergence 
in the exact expressions for the propagators.

In this paper this effect will be taken into account as follows: the propagators will be taken in  
form (\ref{prop1}), but the integration in
the expression for the formfactor $F(s)$ will be done in the region $x,y \in (x_{min},~1)$. 
In other words to get rid of the singularity the cut off parameters $x_{min}$ is introduced.  

In principle, the calculation with propagators (\ref{prop}) is also possible. As it
was noted already, in this case the terms proportional to the $\delta$ play role of the regulator of expression (\ref{I0}). However, 
one can expect that the calculation done in this manner is less accurate and the result must be smaller than it is. 
To understand this one can apply the idea of duality of NRQCD and LC descriptions 
of the hard exclusive nonrelativistic mesons production: {\it if LC expression for the formfactor $F(s)$
is expanded in relative velocities of quark antiquark pairs of the mesons $V$ and $P$, one will get 
NRQCD result for the formfactor. }\footnote{ It should be noted that these is no strict proof of this 
statement. However, one can expect that this statement is indeed true since the amplitude in NRQCD and LC 
can be expended in  series of equivalent operators. Assuming that both theories can describe 
experiment one can expect that these expansions in both theories coincide. } 

At the leading order approximation of NRQCD this idea can be reproduced if one takes infinitely 
narrow approximation for the DAs 
($P_i(x)=V_j(x) \sim \delta (x-1/2),~i=A, P,~~j=T,L,{\perp},A$ ) and uses the following 
values of the parameters in formula (\ref{previous}) $M_V=M_P=2 M_c^*, {\overline M}_c=M_c^*, \mu=M_c^*, f_{V,P,T}^2= \langle O_1 \rangle_S/M_c^*$.
If this procedure is applied to (\ref{previous}) with propagators (\ref{prop1}), 
we will exactly reproduce the expression for the formfactor $F(s)$ obtained within the leading order of 
NRQCD \cite{Braaten:2002fi} 
\beq
F_{NRQCD}=\frac {2^9 \pi \alpha_s} {9 s^2} \langle O_1 \rangle_S,
\label{NR}
\eeq
where $\langle O_1 \rangle_S$ is the NRQCD matrix element \cite{Braaten:2002fi}.
Note that at the leading order approximation of NRQCD  the formfactor $F(s)$
scales exactly as $1/s^2$. So, within NRQCD $1/s^3$ terms appear due to the relativistic corrections,
which are suppressed as $v^2$ ($v^2$ is the characteristic velocity in charmonium). In LC 
$1/s^3$ terms can appear only due to the power corrections of the leading order result. 
So, applying the idea of 
duality of NRQCD and LC one can state that  the power $1/s$ 
corrections to formula (\ref{previous}) with propagators (\ref{prop1}) are 
of order of $\sim v^2 (M_c^*)^2/s$. If we further apply the same 
procedure but with propagators (\ref{prop}), the formfactor 
$F(s)$ will be different from the leading order NRQCD prediction
\beq
F(s)=F_{NRQCD} (1-14 \delta+ O(\delta^2)) = F_{NRQCD} \biggl ( 1-14 
\frac {(M_c^*)^2} {s} + O \biggl (\frac 1 {s^2} \biggr ) \biggr ).
\label{est}
\eeq
To get the agreement with NRQCD prediction for the formfactor, one must expect rather large power correction 
to this result ($14  \delta \times F_{NRQCD}$) in LC so that to cancel the second term in the last equation. 
This correction appears at next-to-leading order approximation in $1/s$ expansion. So, the leading order 
approximation of  LC with propagators (\ref{prop}) underestimates real result. It is not difficult 
to estimate the size of this effect using formula (\ref{est}). For $\sqrt s=10.6$ GeV and $m_c^*=1.4$ GeV the cross section 
calculated with  propagators (\ref{prop}) is smaller than that with propagators (\ref{prop1}) by $\sim 50 \%$. 
Numerical calculation confirms this estimation.

3. To calculate the cross section of the processes $e^+e^- \to J/\Psi \eta_c, J/\Psi \eta_c', \psi' \eta_c, \psi' \eta_c'$ 
one needs to know the following constants:
\beq
\label{const}
\langle J/\Psi (p, \epsilon) | \bar C \gamma_{\alpha} C| 0 \rangle &=& f_{V1} M_{J/\Psi} \epsilon_{\alpha} ~~~~~~~~~~~~~~~~~~~~
\langle \psi'  (p, \epsilon) | \bar C \gamma_{\alpha} C| 0 \rangle = f_{V2} M_{\psi'} \epsilon_{\alpha}, \\ \nonumber
\langle J/\Psi (p, \epsilon) | \bar C \sigma_{\alpha \beta} C| 0 \rangle_{\mu} &=& i f_{T1} (\mu) 
(p_{\alpha} \epsilon_{\beta} - p_{\beta} \epsilon_{\alpha} ) ~~~~ 
\langle \psi'  (p, \epsilon) | \bar C \sigma_{\alpha \beta} C| 0 \rangle_{\mu} =i f_{T2} (\mu)  
(p_{\alpha} \epsilon_{\beta} - p_{\beta} \epsilon_{\alpha} ), \\ \nonumber
\langle \eta_c (p, \epsilon) | \bar C \gamma_{\alpha} \gamma_5 C| 0 \rangle &=& i f_{P1} p_{\alpha}  ~~~~~~~~~~~~~~~~~~~~~~~~~~ 
\langle \eta_c'(p, \epsilon) | \bar C \gamma_{\alpha} \gamma_5 C| 0 \rangle =i f_{P2} p_{\alpha}.   
\eeq

It should be noted that the operator ${\bar C} \sigma_{\alpha \beta}  C$ is not renormalization group invariant. 
For this reason the constants $f_{T1}$ and $f_{T2}$ depend on scale as
\beq
f_{T} (\mu) = \biggl ( \frac {\alpha_s( \mu)} {\alpha_s( \mu_0)} \biggr )^{\frac {c_F} { b_0}} f_{T} (\mu_0),~~~   i=1,2.
\eeq
In the derivation of formula (\ref{previous}) it was assumed  that the tensor and vector constants  $f_{Vi}, f_{Ti}( \overline {M}_c )$ 
are connected to each other as $f_{Ti}( \overline {M}_c )/f_{Vi} = 2\, {M}_c /M_{Vi},~i=1,2$ \cite{Bondar:2004sv, Braguta:2005kr}. At the leading 
order approximation in relative velocity and strong coupling constant these 
relations are correct. However, they are violated due to radiative and relativistic corrections especially for 
the excited mesons. For this reason, below $f_V, f_T$ will be treated as independent constants.

Introducing the modifications described above, the expression for the formfactor $F(s)$ can be rewritten as follows
\beq
|F (s)| & = & \frac{32\pi}{9}   \left|\frac{f_V f_P  M_P M_V}{s^2}\right|\,I_0\,,
\label{res}
\eeq
\beq
\label{I0res}
&&I_0  = 
  \int^1_{x_{min}} dx \int^1_{y_{min}} dy~ \alpha_s(\mu) \left\{
     \frac {\tilde f_t} {M_V} \frac {M_P} {2 {\overline M}_c} \frac{Z_p 
    V_{T}(x, \mu) P_{P}(y, \mu)}{x^2\,y}- \frac {\tilde f_t} {M_P} \frac{\overline {M}_c }{2 M_V}\,
    \frac{Z_m V_T(x, \mu) P_A(y, \mu)}{x^2\,y}+
\right.\\ \nonumber & + &
\frac{1}{2 M_P}\frac{V_{L}(x, \mu)\,P_{A}(y, \mu)}{x \, y}+\frac{(1-2 y)}{2 M_P}
\frac{V_{\perp}(x, \mu)\,P_{A}(y, \mu)}{x \,y^2}+
 \left.
\frac{1}{8} \biggl ( 1- {\tilde f_t} Z_m  \frac{2 {\overline M}_c } { { M_V} }\biggr ) \frac 1 M_P \, 
\frac{(1+y)V_A(x, \mu)P_A(y, \mu)}
{x^2 \,y^2}\right\},
\eeq
where $\tilde f_t = f_T(\mu)/f_V$. Now we are ready to proceed with the calculation. 

\section{Numerical results.}

To calculate the cross sections of the processes $e^+e^- \to J/\Psi \eta_c, J/\Psi \eta_c', \psi' \eta_c, \psi' \eta_c'$  
the following values of input  parameters will be used:

{\bf 1.} The strong coupling constant $\alpha_s ( \mu) $ will be taken at the one loop approximation
\beq
\alpha_s (\mu) = \frac {4 \pi} {\beta_0 \log (\mu^2/\Lambda^2)},
\eeq
with $\Lambda=0.2$ GeV, $\beta_0=25/3$. 

{\bf 2.} For the $\overline {MS}$ mass and the pole mass of $c$-quark the values  
${\overline M}_c=1.2$ GeV and $ M_c^*=1.4$ GeV will be used. 

{\bf 3.} As it was noted in the previous section, to calculate the cross sections one needs 
constants (\ref{const}). The constants $f_{V1}$ and $f_{V2}$ can be determined directly 
from the experiment. The constants $f_{T1}$ and $f_{T2}$ were calculated 
within NRQCD in paper \cite{Braguta:2007ge}. NRQCD formalism can also be used to determine the values of the last
two constants $f_{P1}$ and $f_{P2}$ (see Appendix A). So, the calculation will be done with 
the following values of these constants
\beq
\label{const1}
f_{V1}^2 &=& 0.173 \pm 0.004~ \mbox{GeV}^2, \qquad~~~~~~ f_{V2}^2 = 0.092 \pm 0.002~ \mbox{GeV}^2, \\ \nonumber 
f_{T1}^2(M_{J/\Psi}) &=& 0.144 \pm 0.016~ \mbox{GeV}^2, \quad f_{T2}^2 (M_{J/\Psi}) = 0.068 \pm 0.022~ \mbox{GeV}^2, \\ \nonumber 
f_{P1}^2 &=& 0.139 \pm 0.048~ \mbox{GeV}^2, \quad~~~~~~~~~ f_{P2}^2 = 0.068 \pm 0.040~ \mbox{GeV}^2. 
\eeq

{\bf 4.} The last inputs to formula (\ref{res}) are the DAs. The models of the leading twist DAs
will be taken from papers \cite{Braguta:2006wr, Braguta:2007fh, Braguta:2007tq}. To build
the models for the twist-3 DAs one can apply equations of motion. This procedure is described 
in detail in Appendix B.

\begin{table}
$$\begin{array}{|c|c|c|c|c|c|c|c|c|}
\hline
 H_1 H_2 & \multicolumn{2}{c||} {\sigma_{Exp}\times Br_{H_2 \to charged >2} (\mbox{fb}) } 
   &  \multicolumn{2}{c||} {\sigma_{LO~NRQCD} (\mbox{fb}) }   
 & \multicolumn{2}{c||} {\sigma_{NRQCD} (\mbox{fb}) }    & \sigma_{poten~model} (\mbox{fb}) & \sigma_{LC} (\mbox{fb}) \\
\hline
 &  \mbox{Belle} \mbox{\cite{Abe:2004ww}}  
 & \mbox{BaBar}  \mbox{\cite{Aubert:2005tj}}   
  & \mbox{\cite{Braaten:2002fi}}  & ~~\mbox{\cite{Liu:1}}~~  & \mbox{\cite{He:2007te}} & \mbox{\cite{Bodwin:2007ga}} & \mbox{\cite{Ebert:2008kj}} & \\
\hline
\JP(1S) \ec(1S)&  25.6 \pm 2.8 \pm 3.4  & 17.6 \pm 2.8^{+1.5}_{-2.1}& 3.78 \pm 1.26 &  5.5  & 20.4 & 17.6^{+10.7}_{-8.3} & 22.2 \pm 1.1 & 14.4^{+11.2}_{-9.8}\\
\hline
\JP(2S) \ec(1S)&  16.3 \pm 4.6 \pm 3.9 & - & 1.57 \pm 0.52  & 3.7  & - & - & 15.3 \pm 0.8 & 10.4^{+9.2}_{-7.8} \\
\hline
 \JP(1S) \ec(2S)&  16.5 \pm 3.0 \pm 2.4  & 16.4 \pm 3.7^{+2.4}_{-3.0}& 1.57 \pm 0.52  & 3.7  & -  & - & 16.4 \pm 0.8 & 13.0^{+12.2}_{-11.0} \\
\hline
 \JP(2S) \ec(2S)& 16.0 \pm 5.1 \pm 3.8 & - & 0.65 \pm 0.22  & 2.5  & - & - & 9.6 \pm 0.5 & 9.0^{+9.7}_{-8.5} \\
\hline
\end{array}$$
\label{tab1}
\caption{ The second and third columns 
contain experimental results measured at Belle and Babar experiments. The $Br_{H_2 \to charged>2}$ means 
the branching ratio of the decay of the hadron $H_2$ into two charged particles. 
In the fourth and fifth columns the results of the leading order NRQCD  
obtained in papers \cite{Liu:1, Braaten:2002fi} are shown. The NRQCD results 
obtained with inclusion of radiative and relativistic  corrections \cite{He:2007te, Bodwin:2007ga} are shown in 
columns six and seven. Potential model predictions \cite{Ebert:2008kj} for the cross sections are 
presented in  column eight. Last column contains the values of the 
cross sections obtained in this paper. 
}
\end{table}

There are different sources of uncertainty to the results obtained in this paper. The most important 
uncertainties can be divided into the following groups:

{\bf 1.} {\it The uncertainty in the models of the distribution amplitudes $\phi_i (x, \mu)$}, 
which can be modeled by the variation of the parameters of these models (\ref{mod_lt}).
The calculation shows that for the processes $e^+e^- \to J/\Psi \eta_c, J/\Psi \eta_c', \psi' \eta_c, \psi' \eta_c'$. 
these uncertainties are not greater than $\sim 5 \%,~ 12 \%,~ 28 \%,~ 40 \%$ correspondingly. It is seen
that these uncertainties are not  very large. This fact results from the property
found in papers \cite{Braguta:2006wr}: the evolution improves any model of DA. 

{\bf 2.} {\it The uncertainty due to the radiative corrections}. 
In the approach applied in this paper the leading logarithmic radiative corrections due to the 
evolution of the DAs and strong coupling constant were resummed. As it was noted above for the leading twist processes
the resummation of the leading logarithms in the DAs and strong coupling constant is equivalent to the resummation of the 
leading logarithms in the whole amplitude. Unfortunately, this is not valid for the next-to-leading 
twist processes, for which some of the leading logarithms are lost.  For this reason 
the radiative corrections should be estimated as  $\sim \alpha_s( \sqrt s /2 ) \log (s/4/(M_c^*)^2) \sim 50 \%$. 

{\bf 3.} {\it The uncertainty due to the power corrections.} This uncertainty is determined 
by the next-to-leading order contribution in the $1/s$ expansion. As it was noted above due to the application 
of the propagators in form (\ref{prop1}) one can hope that these corrections are suppressed by 
the square of relative velocity of quark antiquark pair in the meson. In the 
calculation these corrections will be estimated as $\sim 4 v^2_{\psi'} M_{\psi'}^2/s \sim 25 \%$.

{\bf 4.} {\it The uncertainty due to the regularization procedure.} As it was noted 
above, to get rid of the divergence in the formfactor $F(s)$ the cut of parameter was introduced.
Evidently, our results depend on the value of the cut of parameter $x_{min} \cdot \sqrt s$, 
which is of order of the mass of $c$-quark (see formula (\ref{xmin}) ). To estimate this source 
of uncertainty the cut off parameter is varied in the region $x_{min} \cdot \sqrt s =1.0-1.6$ GeV. 
The calculation shows that at $x_{min} \cdot \sqrt s =1.0$ GeV the cross  sections are increased 
by $\sim 50 \%$ and at $x_{min} \cdot \sqrt s =1.6$ GeV the cross  sections are decreased 
by $\sim 20 \%$. It should be noted that this source of uncertainty is closely connected 
with the uncertainty due to the radiative corrections. However, to  understand this in detail 
one needs the theory which takes into account all the leading logarithmic corrections.

{\bf 5.} {\it The uncertainty in the values of constants (\ref{const1}).} It should be noted 
that this source of uncertainty is very important especially for the production 
of the excited states $\psi', \eta_c'$. Thus, for the processes 
$e^+e^- \to J/\Psi \eta_c, \psi' \eta_c, J/\Psi \eta_c', \psi' \eta_c'$ the
errors due to the uncertainties in the values of constants (\ref{const1}) are $\sim 35 \%, 40 \%, 60 \%,  65 \%$
correspondingly. 

Adding all the uncertainties in quadrature one gets the total errors of the calculations.

The results of the calculation are presented in Table I. The second and third columns 
contain experimental results measured at Belle and Babar experiments. 
In the fourth and fifth columns the results of the leading order of NRQCD approach 
obtained in papers \cite{Liu:1, Braaten:2002fi} are shown. The NRQCD results 
obtained with inclusion of radiative and relativistic  corrections \cite{He:2007te, Bodwin:2007ga} are shown in 
columns six and seven. Potential model predictions \cite{Ebert:2008kj} for the cross sections are 
presented in  column eight. Last column contains the values of the 
cross sections obtained in this paper.

\section{Discussion.}

It is seen from Tab. I that within the accuracy 
of the calculation the results of this paper are in agreement with 
 Belle and BaBar experiments. It is also seen  that the uncertainty of the calculation is rather large. 
There are two very important sources of uncertainty. 
The first one is the theoretical problem with 
taking into account of all  leading logarithmic radiative corrections to the 
amplitude of the next-to-leading twist processes. It can be estimated as 
$\sim 50-70 \%$  of the cross sections. This source of uncertainty 
can be reduced if the theory, which takes into account all leading logarithmic corrections,
is created. The second very important source of uncertainty is poor knowledge of 
constants (\ref{const}). For some processes this uncertainty can reach $60 \%$. 
The values of these constants used in this paper 
can be considered as the first  estimation of their real values. So, theoretical and
experimental study of these constants can greatly improve the accuracy of any  predictions
done within LC.

\begin{table}
$$\begin{array}{|c|c|c|c|c|}
\hline
 H_1 H_2 & \sigma_{LO~ NRQCD} (\mbox{fb}) & \sigma_{rel~ corr} (\mbox{fb}) & \sigma_{tot} (\mbox{fb}) \\
\hline
\JP(1S) \ec(1S)  & 1.9   & 6.3 & 14.4  \\
\hline 
\JP(2S) \ec(1S)  & 1.0  &  6.2  & 13.0 \\
\hline
 \JP(1S) \ec(2S) & 1.0   & 7.8 & 10.4 \\
\hline
 \JP(2S) \ec(2S) & 0.53  & 7.2  & 9.0  \\
\hline
\end{array}$$
\caption{The second column contains the values of the cross sections obtained at
the leading NRQCD approximation. The third column contains the cross sections 
calculated in the following approximation: all relativistic corrections are resummed, but the leading logarithmic 
radiative corrections are not taken into account. The last column 
represents the results obtained if the relativistic and leading logarithmic radiative corrections 
to the amplitude are taken into account simultaneously. }
\label{tab2}
\end{table}

Next, let us discuss the results obtained in other papers and compare them with 
the results of this paper. First, let us consider the results of the leading 
order NRQCD predictions \cite{Braaten:2002fi, Liu:1} shown in columns four and five of Tab. I.
These results are approximately by an order of magnitude smaller than the cross sections
measured at the experiments. At the same time LC predictions are in reasonable agreement 
with the experiments. This facts lead to the question: why LC predictions are much 
greater than the leading order NRQCD predictions? The answer to this question can be given 
within LC. As it was noted in section III, LC can reproduce the leading order NRQCD results. 
To do this the renormalization group evolution of the constants and DAs will be disregarded 
and all parameters will be taken at the leading NRQCD approximation: 
the constants $f_{Ti},f_{Pi}$ are equal to the constant $f_{Vi}, i=1,2$, 
which can be determined from the leptonic decay width, the mass $\overline {M}_c=M_c^*=1.4$ GeV,
$M_V=M_P=2 M_c^*$,   all DAs are taken at 
the infinitely narrow approximation $\sim \delta(x-1/2)$. Thus one gets the results 
shown in the second column of Tab. II. At the second step  all parameters 
used in the calculation are taken at their central values, but without renormalization 
group evolution. At this step  
infinite series of the relativistic corrections are resummed, but the leading logarithmic 
radiative corrections are not taken into account (see section II). 
The results are shown in the third column of Tab. II. At the last step,
renormalization group evolution is taken into account and the results are
presented in the last column of Tab. II. Within LC this means that 
the relativistic  and leading logarithmic radiative corrections 
are taken into account simultaneously.

From  Tab. II one sees that the relativistic and leading 
logarithmic radiative corrections taken into account simultaneously dramatically enhance the leading
NRQCD predictions and bring the agreement with Belle and BaBar experiments. Very important conclusion which 
can be drawn from this result is that {\it in  hard exclusive processes
relativistic and leading logarithmic radiative corrections play very important role 
and the consideration of such processes at the leading NRQCD approximation is unreliable. }

Looking to the results of  Tab. II one can also draw a conclusion that relativistic corrections alone cannot describe 
the experimental results. This conclusion is in agreement with the results of papers \cite{Braaten:2002fi, Braguta:2005gw, 
Ebert:2006xq, Berezhnoy:2007sp}. In these papers the cross section of the process $e^+ e^- \to J/\Psi \eta_c$ 
is enhanced due to the relativistic corrections by a factor of $\sim 2-3$ what is in agreement 
with the results obtained in this paper. At the same time this conclusion is in disagreement
with the results of paper  \cite{Ebert:2008kj} (see Tab.I column eight), where the authors 
tried to attribute the disagreement between theory and experiment only to the relativistic corrections, 
which were calculated within potential model. 
From the results shown in Tab.II one sees that this approximation is realistic only for the production 
of excited states $e^+ e^- \to \psi' \eta_c'$. In this case the relativistic 
corrections are much more important than the leading logarithmic radiative corrections. 

The authors of papers \cite{Bondar:2004sv, Braguta:2005kr} took into account 
only the part of the leading logarithmic radiative corrections which appears due to the 
evolution of the constant $f_T$ and the running mass of $c$-quark. The evolution 
of the DAs was disregarded. The calculation done in this paper shows that this approximation 
can be applied only to the DAs of the $2S$ charmonia mesons, for which the evolution 
is not very important ( see paper \cite{Braguta:2007tq}). As it was shown in papers 
\cite{Braguta:2006wr, Braguta:2007fh} the evolution of the $1S$ state charmonia DAs 
is very important. To compensate the effect of the evolution of the DAs the authors of paper 
\cite{Bondar:2004sv} proposed rather wide model of the DAs with so called relativistic 
tail\footnote{It should be noted that relativistic tail of the DAs used in this paper 
is absent at the scale $\mu \sim M_c^*$ and appears due to the evolution of the DAs at larger scales.} 
(see paper \cite{Bodwin:2006dm}). 
For instance, the characteristic velocity of the $1S$ charmonia (see formula (\ref{vel})) with 
this DA can be estimated as 
\beq
\langle v^2 \rangle_{1S} \sim 3 \langle \xi^2 \rangle_{1S} = 0.39,
\eeq
which is much larger than $\langle v^2 \rangle_{1S}=0.2-0.3$ calculated in 
papers \cite{Bodwin:2006dn, Braguta:2006wr, Braguta:2007fh, Choi:2007ze, Bodwin:2007fz}.

At the end of this section let us consider how the disagreement between 
theory and experiment can be resolved within NRQCD. The authors 
of papers \cite{He:2007te, Bodwin:2007ga} took into account the relativistic and 
one loop radiative corrections and got the following values of the cross section
of the process $e^+e^- \to J/\Psi \eta_c$
\beq
\sigma(e^+e^- \to J/\Psi \eta_c) &=& 20.4~\mbox{fb} ~~~~~~~~~~ \mbox{\cite{He:2007te}}, \nonumber \\
\sigma(e^+e^- \to J/\Psi \eta_c) &=& 17.6^{+10.7}_{-8.3} ~\mbox{fb} ~~~~ \mbox{\cite{Bodwin:2007ga}}.
\label{nrqcd}
\eeq
So, similarly to this paper, the authors of papers \cite{He:2007te, Bodwin:2007ga} resolved the 
disagreement through the taking into account of the relativistic and radiative corrections. 

Central values (\ref{nrqcd}) are larger than the  central values of the cross section obtained in this paper. 
Below it will be shown that central values (\ref{nrqcd}) are overestimated. A simple way to understand 
why the cross section was overestimated is to consider paper \cite{Gong:2007db} which has 
similar problem. In this paper explicit expression for the one loop radiative corrections to the cross section
of the process $e^+e^- \to J/\Psi \eta_c$ were calculated. The result of this paper 
can be written in the form
\beq
\sigma=\sigma^0 \times \biggl ( 1 + \frac {\alpha_s(\mu)} {\pi} K \biggr ),
\label{snr}
\eeq
where $\sigma^0$ is the cross section at the leading order approximation, 
the expression for the factor $K$ 
can be found in \cite{Gong:2007db}. The cross section $\sigma^0$ 
is proportional to $|R_{J/\Psi}(0)|^2 \times |R_{\eta_c}(0)|^2$, where $R_{J/\Psi}(r), R_{\eta_c}(r)$ 
are the radial wave functions of the  $J/\Psi$ and $\eta_c$ mesons. It should be 
noted here that the cross section $\sigma$ is very sensitive to the values of the 
wave functions at the origin $R_{J/\Psi}(0), R_{\eta_c}(0)$, so it is very important 
how these parameters were calculated. In the calculation 
the authors of \cite{Gong:2007db} took $|R_{J/\Psi}(0)|=|R_{\eta_c}(0)|$ and the value of the $|R_{J/\Psi}(0)|$ was taken from 
the leptonic decay width $\Gamma_{ee}$
\beq
|R_{J/\Psi}(0)|^2 = \frac {1} {1-\frac {16} {3} \frac {\alpha_s} {\pi}} 
\frac {M_{J/\Psi}^2 \Gamma_{ee}} {4 \alpha^2 q_c^2}.
\label{psi0}
\eeq
Now few comments are in order. First,  this expression is taken at the 
next-to-leading order approximation in the strong coupling constant, what means that 
some part of the one loop radiative corrections is put to the $\sigma^0$. 
Second, if formula (\ref{psi0}) is expanded in the $\alpha_s$, one will get 
infinite series in the strong coupling constant. So, the application of 
formula  (\ref{psi0}) in the calculation of the wave function at the origin  
is equivalent to the statement that one knows all infinite series of the radiative 
corrections to the wave function. Evidently, this is not correct. To be 
consistent with the one loop approximation applied in paper \cite{Gong:2007db},
one should expand expression (\ref{psi0}) in the strong coupling constant 
and than leave only  the first term in this expansion
\beq
|R_{J/\Psi}(0)|^2 = \biggl ( 1+\frac {16} {3} \frac {\alpha_s} {\pi} \biggr ) 
\frac {M_{J/\Psi}^2 \Gamma_{ee}} {4 \alpha^2 q_c^2}.
\label{psi1}
\eeq
It is not difficult to see that the $|R_{J/\Psi}(0)|^2$ calculated from formula (\ref{psi0})
is greater than that calculated form formula (\ref{psi1}) by a factor 
$1/(1-(16 \alpha_s/3 \pi)^2)$. The cross section calculated using formula (\ref{psi0})
is greater than the cross section calculated using formula (\ref{psi1}) by a factor 
$1/(1-(16 \alpha_s/3 \pi)^2)^2 \sim 1.5$ for  $\alpha_s=0.25$. So, the values of the wave functions 
at the origin calculated in paper \cite{Gong:2007db} were overestimated, what led 
to the overestimation of the cross section by a factor of $\sim 1.5$. 

Similar overestimation of the wave functions at the origin and, as the result,
overestimation of the cross section takes place in papers \cite{He:2007te, Bodwin:2007ga}. 
For instance, in paper \cite{Bodwin:2007ga} the authors used the value of the 
NRQCD matrix element $\langle O_1 \rangle_{J/\Psi}$, which is analog of the wave function at the 
origin, calculated in paper \cite{Bodwin:2007fz} using the formula
\beq
\langle O_1 \rangle_{J/\Psi} =  \frac {1} {(1-f(\langle v^2 \rangle_{J/\Psi})-\frac {8} {3} \frac {\alpha_s} {\pi} )^2} 
\frac {3 M_{J/\Psi}^2 \Gamma_{ee}} {8 \pi \alpha^2 q_c^2}, 
\label{nrqcd_o}
\eeq
where $\langle v^2 \rangle_{J/\Psi}$ is defined in (\ref{vel}), $f(x)=x/(3(1+x+\sqrt {1+x}))$.
Similar expression can be written for the $\langle O_1 \rangle_{\eta_c}$. It is clear 
that the application of formula (\ref{nrqcd_o}) is equivalent to the statement that 
one knows not only full $\alpha_s$ corrections in front of the leading NRQCD operator 
$\langle O_1 \rangle_{J/\Psi}$ but also full $\alpha_s$ corrections in front of all 
operators that control relativistic corrections $\langle v^n \rangle_{J/\Psi}$ to 
the amplitude of the leptonic decay. Evidently, this is not correct. To make formula (\ref{nrqcd_o})
more consistent with the approach applied in paper \cite{Bodwin:2007ga}, it should be
modified as follows
\beq
\langle O_1 \rangle_{J/\Psi} = \frac {3 M_{J/\Psi}^2 \Gamma_{ee}} {8 \pi \alpha^2 q_c^2 }
\biggl (
\frac {1} {(1-f(\langle v^2 \rangle_{J/\Psi}))^2  }
+\frac {16} {3} \frac {\alpha_s} {\pi}
\biggr ). 
\eeq
In last formula all problems mentioned above are resolved. Similarly, one can improve 
the expression for the operator 
$\langle O_1 \rangle_{\eta_c}$. With numerical values of paper \cite{Bodwin:2007fz} 
the factor of the overestimation is  $\sim 1.5$. Taking into account this factor,
the cross section $17.6$ fb obtained in paper \cite{Bodwin:2007ga} should be changed to $11.7$ fb.

Similar analysis can be done for the result of paper \cite{He:2007te}. The 
calculation shows that the cross section is overestimated by a factor of $\sim 1.5$.
Taking into account this factor the central value of the cross sections now is $\sim 13.6$ fb.

\section{Conclusion.}

This paper is devoted to the study of the processes 
$e^+e^- \to  J/\Psi \eta_c, J/\Psi \eta_c', \psi' \eta_c, \psi' \eta_c'$ within Light Cone Formalism (LC). 
The amplitude for these processes was first derived in paper \cite{Bondar:2004sv}. 
In the present paper the amplitude was modified so that to archive better accuracy of the 
calculation and to resolve some questions raised in \cite{Bodwin:2006dm}. 

To calculate the cross sections of the processes under study one needs
the twist-2 and twist-3 distribution amplitudes (DA). The models of the twist-2
DAs of the $S$-wave charmonia were proposed in papers \cite{Braguta:2006wr, 
Braguta:2007fh, Braguta:2007tq}. To get the twist-3 DAs, equations of motion 
were applied. It turns out that if one ignores the contribution arising from 
 higher fock states, the twist-3 DAs can be unambiguously determined. 

Using the models of the DAs, the cross sections of the processes 
$e^+e^- \to  J/\Psi \eta_c, J/\Psi \eta_c', \psi' \eta_c, \psi' \eta_c'$ were calculated. 
Within the error of the calculation the results of this study are in agreement
with  Belle and BaBar experiments. In addition, the question - why LC predictions are much 
greater than the leading order NRQCD predictions - was studied. 
Numerical results of the calculation shows that large disagreement 
between LC and the leading NRQCD predictions can be attributed to  
large contribution of  relativistic and radiative corrections. 
From this results one can draw a conclusion that in hard exclusive processes
 relativistic and  radiative corrections play very important role 
and the consideration of such processes at the leading NRQCD approximation 
is unreliable. 

The results of this paper are in agreement with recent NRQCD study 
of the process $e^+e^- \to J/\Psi \eta_c$ \cite{He:2007te, Bodwin:2007ga}  where the authors took into 
account  relativistic and one loop radiative corrections. However, 
in the present paper it was shown that the results of these papers are 
overestimated by a factor 1.5.

\begin{acknowledgments}
The author thanks A.K. Likhoded, A.V. Luchinsky for useful discussion.
This work was partially supported by Russian Foundation of Basic Research under grant 07-02-00417, 
CRDF grant Y3-P-11-05 and president grant MK-2996.2007.2.

\end{acknowledgments}

\appendix
\section{ Calculation of the constants $f_{P1}, f_{P2}$. }

To calculate the values of the constants $f_{Pi},~ i=1,2$  one can apply NRQCD formalism. 
At the NLO approximation of NRQCD the constants $f_{Pi}$ can be written as follows 
\cite{Bodwin:1994jh, Braaten:1998au}
\beq
\label{fp}
f_{Pi}^2 M_{P_i}^2 &=& \langle P_i | \varphi^+ \chi  | 0  \rangle
\langle 0 | \chi^+ \varphi | P_i  \rangle 
\times \biggl ( 1- 4 \frac {\alpha_s } {\pi}- {\langle v^2 \rangle_i }  \biggr ),
\eeq
where 
\beq
\langle v^2 \rangle_i =- \frac 1 {m_c^2} 
\frac {\langle 0 | \chi^+  ({\overset {\leftrightarrow} {\bf D} })^2 \varphi | P_i  \rangle}
 {\langle 0 | \chi^+  \varphi | P_i  \rangle},
 \label{vel}
\eeq
$P_i$ here is the $\eta_c$ meson  if $i=1$ and the $\eta_c'$ meson if $i=2$. To continue the calculation one 
also needs NRQCD expression for the 
decay width $\Gamma \left [ \eta_c \to \gamma \gamma \right ]$:
\beq
&&\Gamma \left [ \eta_c \to \gamma \gamma \right ] = \frac {4 \pi q_c^2 \alpha^2} {M_{\eta_c}^3} 
\langle \eta_c | \varphi^+ \chi  | 0  \rangle
\langle 0 | \chi^+ \varphi | \eta_c  \rangle  
\times \biggl ( 1 +  \frac {\pi^2-20} 3 \frac {\alpha_s } {\pi}- \frac {\langle v^2 \rangle_{\eta_c} } 3  \biggr )  
\eeq

To determine the constant $f_{P1}$ let us express this constant through the width 
$\Gamma \left [ \eta_c \to \gamma \gamma \right ]$\footnote {In the derivation of (\ref{ff}) 
the expression $M_{\eta_c}=2 M_c^* + M_c^* \langle v^2 \rangle_{\eta_c}$ was used.}
\beq
\label{fp1}
f_{P1}^2= \frac {M_c^*} {2 \pi q_c^2 \alpha^2} \Gamma \left [ \eta_c \to \gamma \gamma \right ] \biggl (
1- \frac {\pi^2 - 8 } {3} \frac {\alpha_s} {\pi} - \frac 1 6 \langle v^2 \rangle_{\eta_c}
\biggr ),
\label{ff}
\eeq
where $M_c^*$ is the mole mass of $c$-quark. 
The value of the constant $f_{P1}$  will be calculated with the following set of parameters:
$\Gamma \left [ \eta_c \to \gamma \gamma \right ]=7.2 \pm 0.7 \pm 2.0$ \cite{Yao:2006px}, $\alpha_s=0.25$, 
$\langle v^2 \rangle_{\eta_c}=0.25$ \cite{Bodwin:2006dn}, 
$\langle v^2 \rangle_{\psi'}=0.54$ \cite{Braguta:2007tq}, $M_c^*=1.4\pm 0.2$ GeV. To estimate the  
error of the calculation one should take into account that within NRQCD the constant is double series 
in the relativistic and radiative corrections. At the NNLO approximation one has the relativistic corrections 
$\sim \langle v^2 \rangle^2$, the radiative corrections to the short distance coefficient of the operator
$\langle 0 | \chi^+ (\vec {\sigma} \vec {\epsilon }) \varphi | V_i(\epsilon)  \rangle$ $\sim \alpha_s^2$ and 
the radiative corrections to the short distance coefficient of the  operator 
$\langle 0 | \chi^+ ( \vec {\sigma} \vec {\epsilon } ) ({\overset {\leftrightarrow} {\bf D} })^2 \varphi | V_i(\epsilon)  \rangle$
that can be estimated as 
$\sim \alpha_s \langle v^2 \rangle$. In addition, there is an experimental uncertainty in the measurement of  
$\Gamma \left [ \eta_c \to \gamma \gamma \right ]$ and the uncertainty in the $m_c$. Adding all these uncertainties in quadrature one can 
estimate the error of the calculation. Thus one gets
\beq
f_{P1}^2 = 0.139 \pm 0.048~~ \mbox{GeV}^2.
\label{fp2}
\eeq
Unfortunately, one cannot apply formula (\ref{fp1}) to get the value of the $f_{P2}$. Since today 
only the product $\Gamma \left [ \eta_c' \to \gamma \gamma \right ] \times Br(\eta_c' \to K^0_S K^{\pm} \pi^0) $  
has been measured \cite{Yao:2006px} and there is no model independent way to determine $\Gamma \left [ \eta_c' \to \gamma \gamma \right ]$.

To estimate the value the $f_{P2}$ one could use the value of the constants $f_{V2}$ of $f_{T2}(M_{J/\Psi})$, which 
are equal up to the relativistic correction and radiative corrections. 
In this paper  the constant $f_{T2}(M_{J/\Psi})$ 
will be taken to get the value of $f_{P2}$
\beq
f^2_{P2} = 0.068 \pm 0.040~~ \mbox{GeV}^2.
\eeq

\section{Models for the distribution amplitudes. }

In this section models and renormalization group evolution of the DAs needed in the calculation
will be considered. The DAs of the vector meson $V$ are defined as follows \cite{Bondar:2004sv}
\beq
\langle V( p, \epsilon ) | \bar Q_{\alpha}(z)  Q_{\beta}(-z) | 0 \rangle_{\mu} = \frac {f_V M_V} 4
\int_0^1 dx e^{i (pz) (2 x-1)} \biggl \{ 
\biggl ( \hat {\epsilon} - \hat p \frac {({\epsilon}z)} {(pz)} \biggr ) V_{\perp} (x, \mu) + 
\hat p \frac {({\epsilon}z)} {(pz)} V_L(x, \mu) 
+  \nonumber \\ \frac {f_T(\mu)} {f_V } \frac 1 {M_V}~ \sigma_{\mu \nu} {\epsilon}^{\mu} p^{\nu}~ V_T(x, \mu) + 
\frac 1 8 \biggl ( 1-  \frac {f_T(\mu)} {f_V}  \frac{2 M_c(\mu) } { { M_V} }\biggr )~ 
e_{\mu \nu \sigma \rho} \gamma^{\mu} \gamma_5 {\epsilon}^{\nu} p^{\sigma} z^{\rho}~ V_A(x, \mu)
\biggr \}_{\beta \alpha}.
\label{vec}
\eeq
The DAs of the pseudoscalar meson are defined as 
\beq
\langle P( p ) | \bar Q_{\alpha}(z)  Q_{\beta}(-z) | 0 \rangle_{\mu} =i \frac {f_P M_V} 4
\int_0^1 dx e^{i (pz) (2 x-1)} \biggl \{ 
\frac {\hat p \gamma_5} {M_P} P_A(x,\mu) - \frac {M_P} {2 M_c(\mu)} \gamma_5  P_P (x,\mu)
\biggr \}_{\beta \alpha}.
\label{pseud}
\eeq
where $x$ is the fraction of momentum of the meson $V$ carried by quark,
the constants $f_T(\mu), f_V$ are defined in equations (\ref{const}), 
$M_c(\mu)$ is the running mass of $c$-quark. Expressions (\ref {vec}, \ref {pseud})
are defined at the scale $\mu$. 

The models of the leading twist DAs $V_L(x,\mu), V_T(x,\mu), P_A(x,\mu)$ 
were proposed in papers \cite{Braguta:2006wr, Braguta:2007fh, Braguta:2007tq}. According to these models 
the functions $P_A(x,\mu \sim \overline{M}_c ), V_T (x,\mu \sim \overline{M}_c ) , V_L (x,\mu \sim \overline{M}_c )$ 
are equal to the function $\phi_{1S} (x, \mu \sim \overline{M}_c)$ for the $1S$ state mesons and to the
function $\phi_{2S} (x, \mu \sim \overline{M}_c)$ for the $2S$ state mesons, which have the form
\cite{Chernyak:1983ej}
\beq
\phi_{1S} (x, \mu \sim \overline{M}_c) &\sim& \varphi_{as} (x) \mbox{ Exp} \biggl [ - \frac {b} {4 x (1-x)}  \biggr], \nonumber \\
\phi_{2S} (x, \mu \sim \overline{M}_c) &\sim& \varphi_{as} (x) ( a + (2 x-1)^2 ) \mbox{ Exp} \biggl [ - \frac {b} {4 x (1-x)}  \biggr],
\label{mod_lt}
\eeq
where  $\varphi_{as}(x)=6 x (1-x)$ is the asymptotic function.
For the $1S$ charmonium states the constant $b$ can vary within the interval $3.8 \pm 0.7$. 
For the $2S$ charmonium states the constants $a$ and $b$ can vary within the intervals $0.03^{+0.32}_{-0.03}$ 
and $2.5^{+3.2}_{-0.8}$ correspondingly. The renormalization group evolution of the leading twist DAs is well 
known \cite{Chernyak:1983ej} and it can be written in the form 
\begin{eqnarray}
\phi (x, \mu) = 6 x (1-x) \biggl [  1 + \sum_{n=2,4..} \biggl ( \frac {\alpha_s( \mu)} 
{\alpha_s( \mu_0)} \biggr )^{ \epsilon_n / {b_0}} a_n(\mu_0) C_n^{3/2} ( 2 x-1 ) \biggr ],
\label{wfl}
\end{eqnarray}
where $a_n(\mu_0)$ is the coefficient of the expansion in Gegenbauer polynomials $C_n^{3/2}(z)$
at scale $\mu_0$, the constant $b_0=25/3$, the anomalous dimensions $\epsilon_n$
for the functions $V_L, P_A$ are defined as 
\beq
\epsilon_n &=& \frac 4 3 \biggl ( 1- \frac 2 {(n+1) (n+2)} + 4 \sum_{j=2}^{n+1} \frac 1 j \biggr ), 
\eeq
the anomalous dimension $\epsilon_n$ for the function $V_T$
\beq
\epsilon_n &=& \frac 4 3 \biggl (  4 \sum_{j=2}^{n+1} \frac 1 j \biggr ).
\eeq
Further let us consider the twist-3 DA $P_P(x,\mu)$. It turns out the if one ignores 
higher fock states it is possible to connect the twist-3 DA $P_P(x,\mu)$ to the twist-2 DA $P_A(x,\mu)$
using equations of motion \cite{Ball:1998je}
\beq
\langle \xi^n \rangle_P = \delta_{n0} + \frac {n-1} {n+1} \biggl ( \langle \xi^{n-2} \rangle_P 
- r(\mu) \langle \xi^{n-2} \rangle_A
\biggr ),
\label{eom1}
\eeq
where $\xi=2 x-1$, $\langle \xi^n \rangle_{P,A}$ are the moments of the DAs $P_P$ and $P_A$ correspondingly, 
$r(\mu)=4 M_c(\mu)^2/M_{\eta}^2$. To solve these equations one can expand the $P_P$
in a series of Gegenbauer polynomials $G^{1/2}_n(z)$ \cite{Braun:1989iv}
\beq
P_P(x, \mu) =  \biggl [  1 + \sum_{n=2,4..} b_n(\mu) C_n^{1/2} ( 2 x-1 ) \biggr ].
\label{wfp}
\eeq
Substituting  expressions for the DAs $P_P$ and $P_A$ (\ref{wfl}) and (\ref{wfp})
to  equations of motion (\ref{eom1}), one can solve these equations  recursively and relate 
the coefficients $b_n(\mu)$ to the coefficients $a_n(\mu)$. For instance, for the first tree coefficients
one can get the following formulas
\beq
b_2(\mu) = - \frac 5 2 r(\mu),~~~ b_4(\mu) = - \frac {27} {20} r(\mu) -\frac {81} {10} r~a_2(\mu),~~~
b_6(\mu) = - \frac {13} {14} r(\mu) - \frac {39} {7} r(\mu) a_2(\mu) - \frac {195} {14} r(\mu) a_4(\mu) .
\eeq
It is clear that relativistic motion in the nonrelativistic system must be 
strongly suppressed. The behavior of DA in the end point region $x \sim 0,1$ 
is determined by the relativistic motion. So, the DA of nonrelativistic system 
must be strongly suppressed in the end point region. As it was shown in paper 
\cite{Braguta:2006wr} for the leading twist DAs such suppression can be achieved if there is fine tuning 
of the coefficients $a_n(\mu)$ at the scale $\mu \sim \overline{M}_c$. Similarly, to get the 
suppression in the end  point region for the function $P_P$ one requires the fine tuning in the coefficients $b_n(\overline{M}_c)$, 
which can be achieved through the fine tuning of the constants $r(\overline{M}_c)$ and $a_n(\overline{M}_c)$. 
If we put $r(\overline{M}_c)=4 \overline{M}_c^2/M_{\eta}^2$ the fine tuning for the 
higher moments will be broken what leads to  large relativistic motion in the end point region. 
To avoid this problem it will be assumed that $r(\overline{M}_c)$ is a free parameter, 
which will be adjusted through the requirement that 
the moments of the $P_P(x,\mu \sim \overline{M}_c)$ must be equal to the moments 
of the $P_A(x,\mu \sim \overline{M}_c)$ to the leading order in 
relative velocity expansion. Using the last statement and (\ref{eom1}) one can obtain the expression 
for the $r(\overline{M}_c)$
\beq
r(\overline{M}_c) = 1- 3 \langle \xi^2 \rangle_A +9 (\langle \xi^2 \rangle_A)^2 - 
5 \langle \xi^4 \rangle_A - 27 (\langle \xi^2 \rangle_A)^3 + 30 
\langle \xi^2 \rangle_A \langle \xi^4 \rangle_A - 7 \langle \xi^6 \rangle_A + O(v^8).
\label{rr}
\eeq
All moments in this expression are taken at scale $\mu \sim \overline{M}_c$. 
The exact expression for the $r(\overline{M}_c)$ can be written in the following 
form
\beq
\frac 2 {r(\overline{M}_c)} =  
\int_{-1}^1 d \xi \frac {1+\xi^2} {(1-\xi^2)^2} P_A \biggl ( \frac {1+\xi} {2}, \mu \sim \overline{M}_c \biggr ),
\eeq
where $\xi=2 x-1$. 

The same approach can be applied to the twist-3 DAs $V_A, V_{\perp}$. These 
functions can be expanded in a series of Gegenbauer polynomials \cite{Ball:1998sk}:
\beq
V_{\perp} (x, \mu) =  \biggl [  1 + \sum_{n=2,4..} c_n(\mu) C_n^{1/2} ( 2 x-1 ) \biggr ], \nonumber \\
V_A  (x, \mu) = 6 x (1-x) \biggl [  1 + \sum_{n=2,4..} d_n(\mu) C_n^{3/2} ( 2 x-1 ) \biggr ].
\eeq
The coefficients $c_n(\mu)$ and $d_n(\mu)$ can  be related to the coefficients 
$a_n(\mu)$ of the functions $V_L, V_T$ through the equations 
of motion \cite{Ball:1998sk}
\beq
(n+1 ) \langle \xi^n \rangle_{\perp} =  \langle \xi^n \rangle_{L} + \frac {n(n-1)} {2} (1-\delta(\mu)) 
\langle \xi^{n-2} \rangle_A,  \\ \nonumber
\frac 1 2 (n+2 ) (1-\delta(\mu) ) \langle \xi^n \rangle_A =  \langle \xi^n \rangle_{\perp} -
\delta(\mu) \langle \xi^{n} \rangle_T,
\eeq
where $\langle \xi^n \rangle_{L,T,\perp, A}$ are the moments of the DAs $V_L, V_T, V_{\perp}, V_A$, 
$\delta(\mu)=2 f_T(\mu)/f_V M_c(\mu) /M_V$. Similarly, to the constant $r(\overline{M}_c)$ the value 
of the constant $\delta( \overline{M}_c)$ must be adjusted. We adjust it through the requirement that 
the moments of the DA $V_{\perp}$ must be equal to the moments of the $V_L$ to the leading order 
approximation in relative velocity expansion.  Thus one can get
\beq
\delta(\overline{M}_c) =1- \biggl ( 2 \langle \xi^2 \rangle_L - 6 (\langle \xi^2 \rangle_L )^2 
+ 4 \langle \xi^4 \rangle_L -22 \langle \xi^2 \rangle_L \langle \xi^4 \rangle_L +
18 (\langle \xi^2 \rangle_L)^3+6 \langle \xi^6 \rangle_L \biggr ) + O(v^8).
\eeq
It should be noted that with this value of the $\delta( \overline{M}_c)$ the moments of the 
function $V_A$ are equal to the moments of the $V_L$ to the leading order in 
relative velocity. The exact expression for the $\delta( \overline{M}_c)$ is 
\beq
\delta( \overline{M}_c) = \frac { \int_{-1}^1 \frac {d \xi} {1-\xi^2 } 
V_L \biggl ( \frac {1+\xi} {2}, \mu \sim \overline{M}_c \biggr )} 
{\int_{-1}^1  {d \xi} \frac {1+\xi^2} {(1-\xi^2)^2 } 
V_T \biggl ( \frac {1+\xi} {2}, \mu \sim \overline{M}_c \biggr )}
\eeq


\begin{thebibliography}{**}

%introduction



%\cite{Braaten:2002fi}
\bibitem{Braaten:2002fi}
  E.~Braaten and J.~Lee,
  %``Exclusive double-charmonium production in e+ e- annihilation,''
  Phys.\ Rev.\ D {\bf 67}, 054007 (2003)
  [arXiv:hep-ph/0211085];


\bibitem{Liu:1}
K.~Y.~Liu, Z.~G.~He and K.~T.~Chao,
  %``Problems of double charm production in e+ e- annihilation at s**(1/2) =
  %10.6-GeV. ((V)),''
  Phys.\ Lett.\ B {\bf 557}, 45 (2003)
  [arXiv:hep-ph/0211181];
  
\bibitem{Liu:2}  
K.~Y.~Liu, Z.~G.~He and K.~T.~Chao,
  %``Search for excited charmonium states in e+ e- annihilation at s**(1/2)  =
  %10.6-GeV,''
  arXiv:hep-ph/0408141.


%\cite{Abe:2002rb}
\bibitem{Abe:2002rb}
  K.~Abe {\it et al.}  [Belle Collaboration],
  Phys.\ Rev.\ Lett.\  {\bf 89}, 142001 (2002),
  [arXiv:hep-ex/0205104].




%\cite{Abe:2004ww}
\bibitem{Abe:2004ww}
  K.~Abe {\it et al.}  [Belle Collaboration],
  %``Study of double charmonium production in e+ e- annihilation at s**(1/2)
  %approx. 10.6-GeV,''
  Phys.\ Rev.\ D {\bf 70}, 071102 (2004),
  hep-ex/0407009.
  %%CITATION = HEP-EX 0407009;%%


%\cite{Aubert:2005tj}
\bibitem{Aubert:2005tj}
  B.~Aubert  [BABAR Collaboration],
  %``Measurement of double charmonium production in e+ e- annihilations at
  %s**(1/2) = 10.6-GeV,''
  hep-ex/0506062.
  %%CITATION = HEP-EX 0506062;%%



%\cite{Bodwin:2002fk}
\bibitem{Bodwin:2002fk}
  G.~T.~Bodwin, J.~Lee and E.~Braaten,
  %``e+ e- annihilation into J/psi + J/psi,''
  Phys.\ Rev.\ Lett.\  {\bf 90}, 162001 (2003)
  [arXiv:hep-ph/0212181].
  %%CITATION = PRLTA,90,162001;%%


%\cite{Bodwin:2002kk}
\bibitem{Bodwin:2002kk}
  G.~T.~Bodwin, J.~Lee and E.~Braaten,
  %``Exclusive double-charmonium production from e+ e- annihilation into two
  %virtual photons,''
  Phys.\ Rev.\  D {\bf 67}, 054023 (2003)
  [Erratum-ibid.\  D {\bf 72}, 099904 (2005)]
  [arXiv:hep-ph/0212352].
  %%CITATION = PHRVA,D67,054023;%%


%\cite{Luchinsky:2003yh}
\bibitem{Luchinsky:2003yh}
  A.~V.~Luchinsky,
  %``On double 1-- charmonium production through two-photon e+ e-  annihilation
  %at s**(1/2) = 10.6-GeV,''
  arXiv:hep-ph/0301190.
  %%CITATION = HEP-PH/0301190;%%

%\cite{Zhang:2005ch}
\bibitem{Zhang:2005ch}
  Y.~J.~Zhang, Y.~j.~Gao and K.~T.~Chao,
  %``Next-to-leading order QCD correction to e+ e- --> J/psi + eta/c at
  %s**(1/2) = 10.6-GeV,''
  Phys.\ Rev.\ Lett.\  {\bf 96}, 092001 (2006)
  [arXiv:hep-ph/0506076].
  %%CITATION = HEP-PH 0506076;%%

%\cite{Gong:2007db}
\bibitem{Gong:2007db}
  B.~Gong and J.~X.~Wang,
  %``QCD corrections to J/psi plus eta_c production in e+e- annihilation at
  %sqrt{s}=10.6 GeV,''
  Phys.\ Rev.\  D {\bf 77}, 054028 (2008)
  [arXiv:0712.4220 [hep-ph]].
  %%CITATION = PHRVA,D77,054028;%%
 
%\cite{Zhang:2008gp}
\bibitem{Zhang:2008gp}
  Y.~J.~Zhang, Y.~Q.~Ma and K.~T.~Chao,
  %``Factorization and NLO QCD correction in e^+ e^- \to J/\psi (\psi(2S))+
  %\chi_{c0} at B Factories},''
  arXiv:0802.3655 [hep-ph].
  %%CITATION = ARXIV:0802.3655;%% 



 
%\cite{Ma:2004qf}
\bibitem{Ma:2004qf}
  J.~P. Ma and Z.~G.Si,
  Phys.\ Rev.\ D {\bf 70}, 074007 (2004), [arXiv:hep-ph/0405111]. 



\bibitem{Bondar:2004sv}
  A.~E.~Bondar and V.~L.~Chernyak,
  % ``Is the BELLE result for the cross section sigma(e+ e- --> J/psi +  eta/c) a
  %real difficulty for QCD?,''
  Phys.\ Lett.\ B {\bf 612}, 215 (2005)
  [arXiv:hep-ph/0412335].
  %%CITATION = HEP-PH 0412335;%%

%\cite{Braguta:2005gw}
\bibitem{Braguta:2005gw}
  V.~V.~Braguta, A.~K.~Likhoded and A.~V.~Luchinsky,
  %``Observation potential for chi/b at the Tevatron and LHC,''
  Phys.\ Rev.\  D {\bf 72}, 094018 (2005)
  [arXiv:hep-ph/0506009].
  %%CITATION = PHRVA,D72,094018;%%


%\cite{Braguta:2005kr}
\bibitem{Braguta:2005kr}
  V.~V.~Braguta, A.~K.~Likhoded and A.~V.~Luchinsky,
   %``Excited charmonium mesons production in e+ e- annihilation at s**(1/2) =
  %10.6-GeV,''
  Phys.\ Rev.\ D {\bf 72}, 074019 (2005)
  [arXiv:hep-ph/0507275].
  %%CITATION = HEP-PH 0507275;%%
  
%\cite{Braguta:2006nf}
\bibitem{Braguta:2006nf}
  V.~V.~Braguta, A.~K.~Likhoded and A.~V.~Luchinsky,
  %``The processes e+ e- --> J/psi chi/c0, psi(2S) chi/c0 at s**(1/2) =
  %10.6-GeV in the framework of light cone formalism,''
  Phys.\ Lett.\  B {\bf 635}, 299 (2006)
  [arXiv:hep-ph/0602047].
  %%CITATION = PHLTA,B635,299;%%


%\cite{Ebert:2006xq}
\bibitem{Ebert:2006xq}
  D.~Ebert and A.~P.~Martynenko,
  %``Relativistic effects in the production of pseudoscalar and vector doubly
  %heavy mesons from e+ e- annihilation,''
  Phys.\ Rev.\  D {\bf 74}, 054008 (2006)
  [arXiv:hep-ph/0605230].
  %%CITATION = PHRVA,D74,054008;%%  



%\cite{Choi:2007ze}
\bibitem{Choi:2007ze}
  H.~M.~Choi and C.~R.~Ji,
  %``Perturbative QCD analysis of exclusive $J/\psi+\eta_c$ production in
  %$e^+e^-$ annihilation,''
  Phys.\ Rev.\  D {\bf 76}, 094010 (2007)
  [arXiv:0707.1173 [hep-ph]].
  %%CITATION = PHRVA,D76,094010;%%

%\cite{Berezhnoy:2007sp}
\bibitem{Berezhnoy:2007sp}
  A.~V.~Berezhnoy,
  %``Comments on the internal motion of massive charm quark in the process of
  %double charmonium production in e+ e- annihilations,''
  arXiv:hep-ph/0703143.
  %%CITATION = HEP-PH/0703143;%%

%\cite{Ebert:2008kj}
\bibitem{Ebert:2008kj}
  D.~Ebert, R.~N.~Faustov, V.~O.~Galkin and A.~P.~Martynenko,
  %``Relativistic description of the double charmonium production in $e^+e^-$
  %annihilation,''
  arXiv:0803.2124 [hep-ph].
  %%CITATION = ARXIV:0803.2124;%%




%\cite{Lee:2003db}
\bibitem{Lee:2003db}
  J.~Lee,
  %``Exclusive two-charmonium vs. charmonium-glueball production at BELLE,''
  J.\ Korean Phys.\ Soc.\  {\bf 45}, S354 (2004)
  [arXiv:hep-ph/0312251].
  %%CITATION = HEP-PH 0312251;%%

%\cite{Zhang:2008ab}
\bibitem{Zhang:2008ab}
  Y.~J.~Zhang, Q.~Zhao and C.~F.~Qiao,
  %``Possible contributions to $e^+e^- \to J/\psi + \eta_c$ due to intermediate
  %meson rescatterings,''
  arXiv:0806.3140 [hep-ph].
  %%CITATION = ARXIV:0806.3140;%%

%\cite{Berezhnoy:2006mz}
\bibitem{Berezhnoy:2006mz}
  A.~V.~Berezhnoy and A.~K.~Likhoded,
  %``Quark-hadron duality and production of charmonia and doubly charmed
  %baryons in e+ e- annihilation,''
  Phys.\ Atom.\ Nucl.\  {\bf 70}, 478 (2007)
  [arXiv:hep-ph/0602041].
  %%CITATION = PANUE,70,478;%%
 
%\cite{Bodwin:1994jh}
\bibitem{Bodwin:1994jh}
  G.~T.~Bodwin, E.~Braaten and G.~P.~Lepage,
  % ``Rigorous QCD analysis of inclusive annihilation and production of heavy
  %quarkonium,''
  Phys.\ Rev.\ D {\bf 51}, 1125 (1995)
  [Erratum-ibid.\ D {\bf 55}, 5853 (1997)]
  [arXiv:hep-ph/9407339].
  %%CITATION = HEP-PH 9407339;%%
 
 
 
 
 
%\cite{He:2007te}
\bibitem{He:2007te}
  Z.~G.~He, Y.~Fan and K.~T.~Chao,
  %``Relativistic corrections to $J/\psi$ exclusive and inclusive double   charm
  %production at B factories,''
  Phys.\ Rev.\  D {\bf 75}, 074011 (2007)
  [arXiv:hep-ph/0702239].
  %%CITATION = PHRVA,D75,074011;%%



%\cite{Bodwin:2007ga}
\bibitem{Bodwin:2007ga}
  G.~T.~Bodwin, J.~Lee and C.~Yu,
  %``Resummation of Relativistic Corrections to e+ e- -> J/psi+eta_c,''
  Phys.\ Rev.\  D {\bf 77}, 094018 (2008)
  [arXiv:0710.0995 [hep-ph]].
  %%CITATION = PHRVA,D77,094018;%%



\bibitem{Bodwin}
G.T. Bodwin, Talk presented at the conference 
Quark Confinement and Hadron Spectrum, Mainz (Germany), 1-6 September 2008.



%\cite{Lepage:1980fj}
\bibitem{Lepage:1980fj}
  G.~P.~Lepage and S.~J.~Brodsky,
  %``Exclusive Processes In Perturbative Quantum Chromodynamics,''
  Phys.\ Rev.\ D {\bf 22}, 2157 (1980).
  %%CITATION = PHRVA,D22,2157;%%

%\cite{Chernyak:1983ej}
\bibitem{Chernyak:1983ej}
  V.~L.~Chernyak and A.~R.~Zhitnitsky,
  %``Asymptotic Behavior Of Exclusive Processes In QCD,''
  Phys.\ Rept.\  {\bf 112}, 173 (1984).
  %%CITATION = PRPLC,112,173;%% 





%\cite{Bodwin:2006dm}
\bibitem{Bodwin:2006dm}
  G.~T.~Bodwin, D.~Kang and J.~Lee,
  %``Reconciling the light-cone and NRQCD approaches to calculating e+ e- -->
  %J/psi + eta/c,''
  Phys.\ Rev.\  D {\bf 74}, 114028 (2006)
  [arXiv:hep-ph/0603185].
  %%CITATION = PHRVA,D74,114028;%%


%\cite{Ma:2006hc}
\bibitem{Ma:2006hc}
  J.~P.~Ma and Z.~G.~Si,
  %``NRQCD factorization for twist-2 light-cone wave-functions of charmonia,''
  Phys.\ Lett.\  B {\bf 647}, 419 (2007)
  [arXiv:hep-ph/0608221].
  %%CITATION = PHLTA,B647,419;%%




%\cite{Braguta:2006wr}
\bibitem{Braguta:2006wr}
  V.~V.~Braguta, A.~K.~Likhoded and A.~V.~Luchinsky,
  %``The study of leading twist light cone wave function of eta/c meson,''
  Phys.\ Lett.\  B {\bf 646}, 80 (2007)
  [arXiv:hep-ph/0611021].
  %%CITATION = PHLTA,B646,80;%%

%\cite{Braguta:2007fh}
\bibitem{Braguta:2007fh}
  V.~V.~Braguta,
  %``The study of leading twist light cone wave functions of J/psi meson,''
  Phys.\ Rev.\  D {\bf 75}, 094016 (2007)
  [arXiv:hep-ph/0701234].
  %%CITATION = PHRVA,D75,094016;%%



%\cite{Braguta:2007tq}
\bibitem{Braguta:2007tq}
  V.~V.~Braguta,
  %``The study of leading twist light cone wave functions of 2S state charmonium
  %mesons,''
  arXiv:0709.3885 [hep-ph].
  %%CITATION = ARXIV:0709.3885;%%


%\cite{Feldmann:2007id}
\bibitem{Feldmann:2007id}
  T.~Feldmann and G.~Bell,
  %``Light-Cone Distribution Amplitudes for Non-Relativistic Bound States,''
  arXiv:0711.4014 [hep-ph].
  %%CITATION = ARXIV:0711.4014;%%

%\cite{Bell:2008er}
\bibitem{Bell:2008er}
  G.~Bell and T.~Feldmann,
  %``Modelling light-cone distribution amplitudes from non-relativistic bound
  %states,''
  JHEP {\bf 0804}, 061 (2008)
  [arXiv:0802.2221 [hep-ph]].
  %%CITATION = JHEPA,0804,061;%%



%\cite{Hwang:2008qi}
\bibitem{Hwang:2008qi}
  C.~W.~Hwang,
  %``Study of quark distribution amplitudes of 1S and 2S heavy quarkonium
  %states,''
  arXiv:0811.0648 [hep-ph].
  %%CITATION = ARXIV:0811.0648;%%
 
%%%%%%%%%%%%%%%%%%%%%%%%%%%%%%%%%%%%%%%%%%%%%%%%% 

%\cite{Braun:2003rp}
\bibitem{Braun:2003rp}
  V.~M.~Braun, G.~P.~Korchemsky and D.~Mueller,
  %``The uses of conformal symmetry in QCD,''
  Prog.\ Part.\ Nucl.\ Phys.\  {\bf 51}, 311 (2003)
  [arXiv:hep-ph/0306057].
  %%CITATION = PPNPD,51,311;%%


%the amplitude of e^+e^- \to J/\Psi \eta_c




%numerical results

%\cite{Braguta:2007ge}
\bibitem{Braguta:2007ge}
  V.~V.~Braguta,
  %``The study of double vector charmonium mesons production at B-factories
  %within light cone formalism,''
  Phys.\ Rev.\  D {\bf 78}, 054025 (2008)
  [arXiv:0712.1475 [hep-ph]].
  %%CITATION = PHRVA,D78,054025;%%

%discussion

%\cite{Bodwin:2006dn}
\bibitem{Bodwin:2006dn}
  G.~T.~Bodwin, D.~Kang and J.~Lee,
  %``Potential-model calculation of an order-v**2 NRQCD matrix element,''
  Phys.\ Rev.\  D {\bf 74}, 014014 (2006)
  [arXiv:hep-ph/0603186].
  %%CITATION = PHRVA,D74,014014;%%
  
%\cite{Bodwin:2007fz}
\bibitem{Bodwin:2007fz}
  G.~T.~Bodwin, H.~S.~Chung, D.~Kang, J.~Lee and C.~Yu,
  %``Improved determination of color-singlet nonrelativistic QCD matrix elements
  %for S-wave charmonium,''
  Phys.\ Rev.\  D {\bf 77}, 094017 (2008)
  [arXiv:0710.0994 [hep-ph]].
  %%CITATION = PHRVA,D77,094017;%%
%appendix


%\cite{Braaten:1998au}
\bibitem{Braaten:1998au}
  E.~Braaten and Y.~Q.~Chen,
  %``Renormalons in electromagnetic annihilation decays of quarkonium,''
  Phys.\ Rev.\  D {\bf 57}, 4236 (1998)
  [Erratum-ibid.\  D {\bf 59}, 079901 (1999)]
  [arXiv:hep-ph/9710357].
  %%CITATION = PHRVA,D57,4236;%%

%\cite{Yao:2006px}
\bibitem{Yao:2006px}
  W.~M.~Yao {\it et al.}  [Particle Data Group],
  %``Review of particle physics,''
  J.\ Phys.\ G {\bf 33}, 1 (2006).
  %%CITATION = JPHGB,G33,1;%%







%-------------------------------


%\cite{Ball:1998je}
\bibitem{Ball:1998je}
  P.~Ball,
  %``Theoretical update of pseudoscalar meson distribution amplitudes of  higher
  %twist: The nonsinglet case,''
  JHEP {\bf 9901}, 010 (1999)
  [arXiv:hep-ph/9812375].
  %%CITATION = JHEPA,9901,010;%%


%\cite{Braun:1989iv}
\bibitem{Braun:1989iv}
  V.~M.~Braun and I.~E.~Filyanov,
  %``Conformal Invariance And Pion Wave Functions Of Nonleading Twist,''
  Z.\ Phys.\  C {\bf 48}, 239 (1990)
  [Sov.\ J.\ Nucl.\ Phys.\  {\bf 52}, 126 (1990\ YAFIA,52,199-213.1990)].
  %%CITATION = YAFIA,52,199;%%



%\cite{Ball:1998sk}
\bibitem{Ball:1998sk}
  P.~Ball, V.~M.~Braun, Y.~Koike and K.~Tanaka,
  %``Higher twist distribution amplitudes of vector mesons in {QCD}: Formalism
  %and twist three distributions,''
  Nucl.\ Phys.\  B {\bf 529}, 323 (1998)
  [arXiv:hep-ph/9802299].
  %%CITATION = NUPHA,B529,323;%%







 




\end{thebibliography}
\end{document}